\documentclass[manuscript]{aastex62}

\submitjournal{ApJ}
\usepackage{lineno}
\usepackage{comment}
\usepackage{amsmath}
\usepackage{amssymb}

\begin{document}

\title{Magnetic Energy Transfer and Distribution between Protons and Electrons for Alfv\'enic Waves at Kinetic Scales in Wavenumber Space}

\correspondingauthor{Jiansen He}
\email{jshept@pku.edu.cn}

\author{Die Duan}
\affiliation{School of Earth and Space Sciences, Peking University \\
Beijing, 100871, China; E-mail: jshept@pku.edu.cn}

\author{Jiansen He}
\affiliation{School of Earth and Space Sciences, Peking University \\
Beijing, 100871, China; E-mail: jshept@pku.edu.cn}

\author{Honghong Wu}
\affiliation{School of Earth and Space Sciences, Peking University \\
Beijing, 100871, China; E-mail: jshept@pku.edu.cn}

\author{Daniel Verscharen}
\affiliation{Mullard Space Science Laboratory, University College London, \\
Holmbury Hill Rd, Dorking RH5 6NT, UK}
\affiliation{Space Science Center, University of New Hampshire, \\
Durham NH 03824, USA}

\begin{abstract}

Turbulent dissipation is considered a main source of heating and acceleration in cosmological plasmas. The alternating current Joule-like term, $\langle\delta\mathbf{j} \cdot \delta\mathbf{E}\rangle$, is used to measure the energy transfer between electromagnetic (EM) fields and particles. Because the electric field depends on the reference frame, in which frame to calculate $\langle\delta\mathbf{j}\cdot \delta\mathbf{E}\rangle$ is an important issue. We compute the scale-dependent energy transfer rate spectrum in wavevector space, and investigate the electric-field fluctuations in two references frames: $\delta \mathbf{E}$ in the mean bulk flow frame and $\delta \mathbf{E}'$ in the local bulk flow frame (non-inertial reference frame). Considering Alfv\'enic waves, we find that $\langle\delta\mathbf{j}\cdot\delta\mathbf{E}^\prime\rangle$, which neglects the contribution of work done by the ion inertial force, is not consistent with the magnetic field energy damping rate ($2\gamma \delta B^2$) according to linear Maxwell-Vlasov theory, while $\langle\delta\mathbf{j}\cdot \delta\mathbf{E}\rangle$  is exactly the same as $2\gamma \delta B^2$ in wavenumber space $(k_\parallel, k_\perp)$, where $\gamma$ is the linear damping rate. Under typical conditions of solar wind at 1 au, we find in our theoretical calculation that the field energy is mainly converted into proton kinetic energy leaving the residual minor portion for electrons. Although the electrons gain energy in the direction perpendicular to the mean magnetic field, they return a significant fraction of their kinetic energy in the parallel direction. Magnetic-field fluctuations can transfer particle energy between the parallel and perpendicular degrees of freedom. Therefore, $\langle\delta\mathbf{j}_\parallel\cdot \delta\mathbf{E}_\parallel\rangle$ and $\langle\delta\mathbf{j}_\perp\cdot \delta\mathbf{E}_\perp\rangle$ cannot solely describe the energy transfer in parallel direction and perpendicular direction, respectively.
\end{abstract}


\section{Introduction}
	\label{S-Introduction}

Turbulence dissipation is an important process in heating and acceleration of particles in extended stellar atmosphere, astrospheres, and the galactic interstellar space. Also the heating of the solar wind is attributed to the dissipation of turbulence. Turbulent dissipation refers to the conversion of turbulent energy into thermal energy or the production of superthermal particle distributions. However, the mechanism of this conversion is not fully understood. In interplanetary turbulence, energy is injected at large scales, then cascades to small scales, and dissipates at even smaller (kinetic) scales \citep{Kiyani2015}. Because of the low density of space plasmas, the dissipation always occurs at scales much smaller than the collisional mean free path of the particles. Therefore, collisionless mechanisms play a vital role in the dissipation \citep{Matthaeus2015, Chen2016, Howes2017}.

Resonant damping is suggested as a mechanism for collisionless dissipation. This kind of resonant interaction between particles and electromagnetic (EM) waves in the plasma, includes Landau damping, transit-time damping and cyclotron-resonant damping \citep{Isenberg1983, Leamon1998, Leamon1999, Gary1999, Marsch2001, Isenberg2001, Klein2017}. Observational evidence for Landau damping and cyclotron damping has been reported recently \citep{He2015a, He2015}. For non-resonant damping, dissipation in coherent structures, such as current sheets \citep{Dmitruk2004, Osman2012}, discontinuities\citep{Wang2013, Zhang2015}, and magnetic reconnection at kinetic scales\citep{Drake2006, Osman2014}, are found both in simulations and observations.  Based on the studies by \citet{Chenliu2001} and \citet{Chandran2010}, the stochastic heating of protons is another effective non-resonant mechanism. All of these mechanisms represent an energy transfer from EM fields to particles, accelerating particles and heating the plasma. In the Vlasov description, this energy transfer corresponds to a change in the particle phase-space density.

The strength of dissipation can be measured by the amount of energy transferred from waves to particles per unit time. This energy transfer is represented by the Joule-like heating term $\mathbf{j}\cdot \mathbf{E}$ ($\mathbf{j}$ is the current density and $\mathbf{E}$ is the electric field) that describes the amount of particle energy gained from the waves per unit time\citep{Stix1992}. The value of $\mathbf{j}\cdot \mathbf{E}$ depends on the reference frame, in which the $\mathbf{E}$ is evaluated. \citet{Zenitani2011} argued that $\mathbf{j}\cdot \mathbf{E}'$, where $\mathbf{E}'$ is the electric field calculated in the local electron bulk flow frame, represents the "true dissipation".  \citet{Wan2015} found that regions of high current density usually have high $\mathbf{j}\cdot \mathbf{E}'$ in their 3D plasma turbulence simulation. \citet{Birn2010} further argued that the plasma heating is contributed by both $\mathbf{j}\cdot\mathbf{E}'$ and  the work of the pressure gradient force ($ \mathbf{v}\cdot(\nabla \cdot\mathbf{P})$) in the energy equation. The particle-in-cell simulation of \citet{Yang2017} demonstrated the importance a different term, $(\mathbf{P}\cdot\nabla)\cdot\mathbf{v}$ for plasma heating. 

In the fast solar wind, Alfv\'enic fluctuations dominate the magnetohydrodynamic (MHD) scales\citep{Belcher1971}. At kinetic scales, turbulence may consist of fluctuations that behave like kinetic Alfv\'en waves, Alfv\'en cyclotron waves, or whistler waves\citep{Galtier2003, Bale2005a, Sahraoui2009, Schekochihin2009, He2012, Gary2012}. In addition, the distribution of turbulent energy is anisotropic with $k_\parallel \ll k_\perp$ in wavenumber space (($k_\parallel, k_\perp$) space)\citep{Goldreich1995, Horbury2008, Podesta2009, Chen2010}. \citet{Narita2010} and \citet{Sahraoui2010} showed the anisotropy of the power spectral density (PSD) ($k_\parallel, k_\perp$) around the ion kinetic range by applying the k-filtering method to Cluster data. \citet{He2013} developed a tomography method to reconstruct the multi-dimensional PSD of magnetic field from data of Helios 2, which reveals an oblique ridge of PSD closer to the $k_\perp$ axis than to the $k_\parallel$ axis. \citet{Yan2016} employed the same method as \citet{He2013} and discovered the anisotropy of the residual energy $E_r=E_v-E_b$ with $E_v=\delta v^2$ and $E_b=\delta b^2$, which is distributed along the $k_\perp$ axis and concentrates at very small $k_\parallel$. The anisotropy of turbulence energy in wavenumber space may be caused by the cascade of Alfv\'en waves preferentially in the perpendicular direction, or by intermittency \citep{Wang2014, Pei2016}. Most previous turbulence studies focus on the analysis of magnetic-energy spectra, yet the  EM energy-conversion-rate spectra have been scarcely investigated and remain unknown. \citet{He2019} measured the EM energy conversion rate spectra in the magnetosheath turbulence, and found that it was enhanced around the ion kinetic scale. On the other hand, the EM energy-conversion rate can also be used to identify the wave excitation and growth of which is a prevalent phenomenon in the fore shock region \citep{He2019b}. In this work, we theoretically predict the energy-conversion-rate spectra around the proton kinetic range for Alfv\'enic waves, and compare these spectra between different reference frames. 

In Section 2, we present our theoretical calculation of the distribution of energy transfer between magnetic field and particle kinetic energy in different reference frames. Section 3 compares the transferred energy partition between protons and electrons in both parallel and perpendicular directions. We discuss the interpretation and implications of our work in Section 4.

\section{Energy Transfer of Alfv\'enic Modes in kinetic theory}
 \label{Section2}
We assume the plasma to consist of only protons and electrons, without background electric field and bulk flow velocity. Both species of particles are isotropic and Maxwellian without drifts. We assume $m_p/m_e = 1836$, $\beta_{p\parallel} = \beta_{p\perp} = \beta_{e\parallel} = \beta_{e\perp} = 1$, and $v_A/c = 0.00016$. We adopt the numerical New Hampshire Dispersion Relation Solver (NHDS) code \citep{Verscharen2018} to calculate the dispersion and polarization relations of wave modes in wavenumber space based on the linearized set of the Vlasov-Maxwell equations. We take the background magnetic field $\mathbf{B}_0 $ along the z direction, and the wavevector $\mathbf{k}$ to be in the x-z plane ($\mathbf{k} = (k_\perp, 0, k_\parallel)$). The frequency of waves is normalized to the proton gyrofrequency $\Omega_p = eB_0/m_p$, and fields are scaled to $\delta B_y$. Around the ion scale ($k\rho_p \sim 1, \rho_p = v_{th,p}/\Omega_p$ is the proton thermal gyro-radius), the Alfv\'enic mode transitions into the kinetic Alfv\'en wave (KAW) for quasi-perpendicular propagation or the ion cyclotron wave (ICW) for quasi-parallel propagation.

\begin{figure}[htb!]
	\centerline{\includegraphics[width=15cm,clip=]{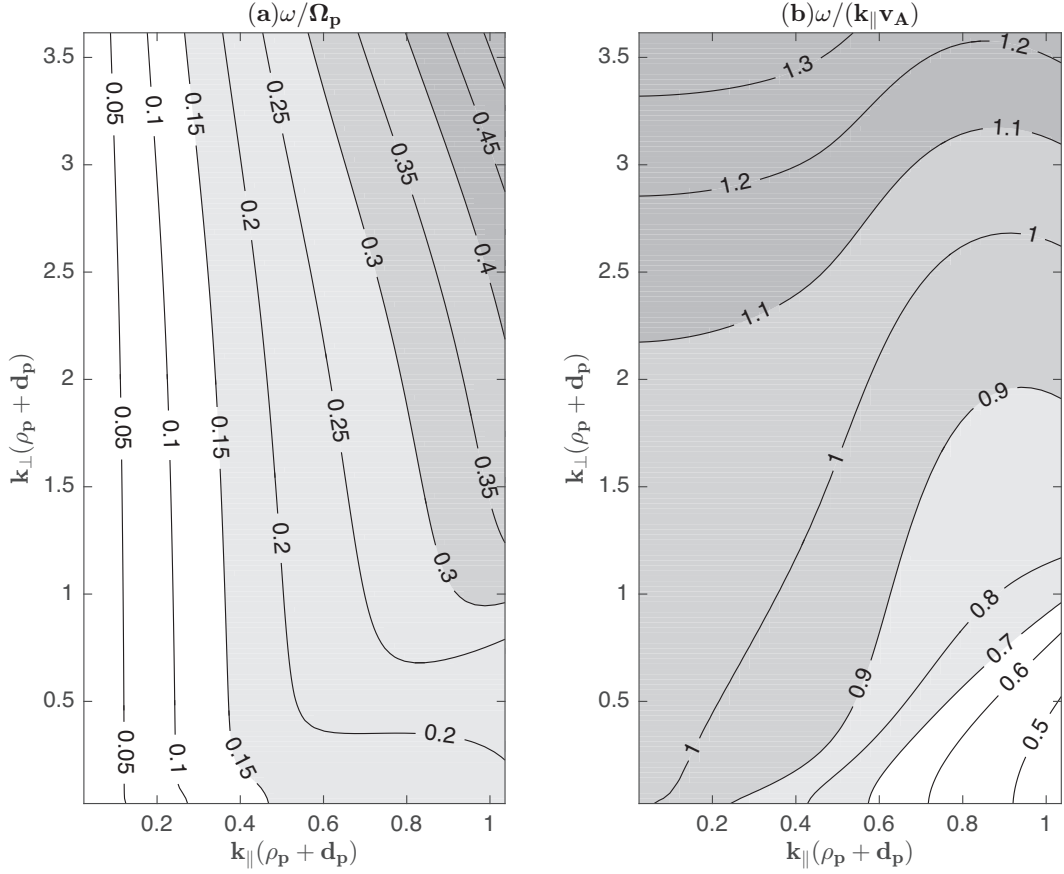}}
	\caption{The dispersion relation of the Alfv\'enic wave mode in wavevector space $(k_\parallel, k_\perp)$ as derived from linear Vlasov-Maxwell theory. (a) The real frequency normalized by the proton gyrofrequency. (b) The parallel phase speed of the waves normalized by the Alfv\'en speed.}
	\label{fig1}
\end{figure}

\begin{figure}[htb!]
	\centerline{\includegraphics[width=13cm, clip=]{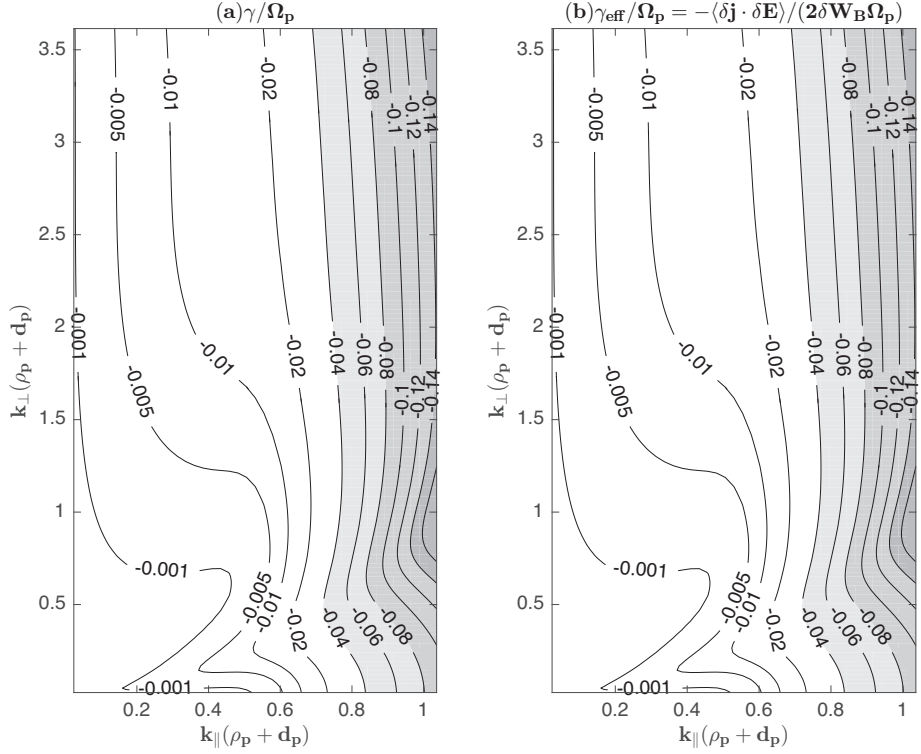}}
	\caption{(a) Damping rate (imaginary frequency normalized to $\Omega_p$) of the Alfv\'enic wave mode. (b) The total effective damping rate due to conversion of EM-field energy to particle kinetic energy. The distributions of the both damping rates are indistinguishable suggesting that the damped magnetic field energy fully converts to particle kinetic energy.}
	\label{fig2}
\end{figure}

From the second moment of the Vlasov equation ($W_s=m_s/2\iiint{v^2f_sdv_xdv_ydv_z}$ is the total kinetic energy and $\mathbf{Q}_s=m_s/2\iiint{v^2\mathbf{v}f_sdv_xdv_ydv_z}$ is the total kinetic energy flux vector), we obtain:
\begin{eqnarray}
	\frac{\partial W_s}{\partial t} + \nabla \cdot \mathbf{Q}_s = \mathbf{j}_s\cdot\mathbf{E}, \label{eq1}
\end{eqnarray}
where the index s=p represents protons and the index s=e represents electrons. Particles gain energy from the the electromagnetic field through the $\mathbf{j}\cdot \mathbf{E}$ term ($\mathbf{j} = \mathbf{j}_p + \mathbf{j}_e$).
We use the distribution of $\delta\mathbf{j}$ and $\delta\mathbf{E}$ in wavenumber space to build $\delta\mathbf{j}\cdot \delta\mathbf{E}$ spectra ($\delta$ represents the fluctuating part of a quantity). The average energy transfer rate over a few periods is given by \citep{Stix1992}: 
\begin{eqnarray} \label{Eq1}
	\langle\delta\mathbf{j}\cdot\delta\mathbf{E}\rangle = \frac{1}{4}(\delta\mathbf{j}^*\cdot\delta\mathbf{E}+\delta\mathbf{j}\cdot\delta\mathbf{E}^*),
\end{eqnarray}
 where the asterisk indicates the complex conjugate. Note that $\delta \mathbf{E}$ and $\delta\mathbf{E}^*$ in Equation \ref{Eq1} are the Fourier amplitudes of the electric field in the plasma frame (or mean bulk flow reference frame), which is an inertial reference frame. When transforming into the local bulk flow reference frame, the electric field can be expressed as $\delta\mathbf{E}^\prime = \delta\mathbf{E} + \delta\mathbf{v} \times (\mathbf{B}_0+\delta\mathbf{B}_0)$, thus the additional inertial forces arise.
In the solar wind, the EM energy is dominated by the energy of the magnetic-field fluctuations. Using the damping rate $\gamma$ (the imaginary part of the wave frequency), we write the magnetic energy damping rate as:
\begin{eqnarray}
	\frac{d\delta W_B}{dt} = 2\gamma\frac{\delta B^2}{2\mu_0} = 2 \gamma \delta W_B,
\end{eqnarray}
where $\delta W_B=\delta B^2/(2\mu_0)$ is the energy of the fluctuating magnetic field. 
The dispersion relation of the Alfv\'en wave branch is shown in Figure \ref{fig1}. The real part of the frequency, $\omega$, increases with $k_\parallel$. $\omega/k_\parallel$ increases with $k_\perp$ as expected. The dispersion relations at $(k_\parallel(\rho_p+d_p)\sim 1,k_\perp(\rho_p+d_p)\sim 0)$ and $(k_\parallel(\rho_p+d_p)\sim 0,k_\perp(\rho_p+d_p)>1)$ represent the characteristics of ICWs and KAWs, respectively. We define the effective damping rate:
\begin{eqnarray}
	\gamma_{\rm{eff}} = - \frac{\langle\delta\mathbf{j}\cdot \delta\mathbf{E}\rangle}{2\delta W_B},
\end{eqnarray}
which describes the ratio of $\langle\delta\mathbf{j}\cdot \delta\mathbf{E}\rangle$ to the fluctuating magnetic field energy. If $\gamma_{\mathrm{eff}}<0$, the EM energy is converted to particle kinetic energy; If $\gamma_{\mathrm{eff}}>0$, the EM fields receive energy from the particles. Figure \ref{fig2} shows that the effective damping rate is equal to the wave damping rate. The coordinates in the figures are scaled as $k(\rho_p + d_p)$, where $d_p = v_A/\Omega_p$ is the proton inertial length. For the case of $k=k_\parallel$ ,this scale refers to the proton cyclotron resonance \citep{Leamon1998}, and this resonance may contribute to the break between the inertial range and dissipation range in the magnetic field PSD of solar wind turbulence \citep{Wang2018, Woodham2018, Duan2018}. The behavior of the effective damping rate illustrates that the fluctuating magnetic energy fully converts to particle kinetic energy through the $ \langle\delta\mathbf{j}\cdot \delta\mathbf{E}\rangle$ term. It also indicates the validity of using $\delta\mathbf{j}$ and $\delta\mathbf{E} $ to estimate the spectrum of energy conversion. The magnetic field energy damps quickly around $k_\parallel(\rho_p + d_p) = 1$, with the normalized damping rate $\gamma/\Omega_p$ approaching -0.1.

\begin{figure}[htb!]
	\centerline{\includegraphics[width=15cm,clip=]{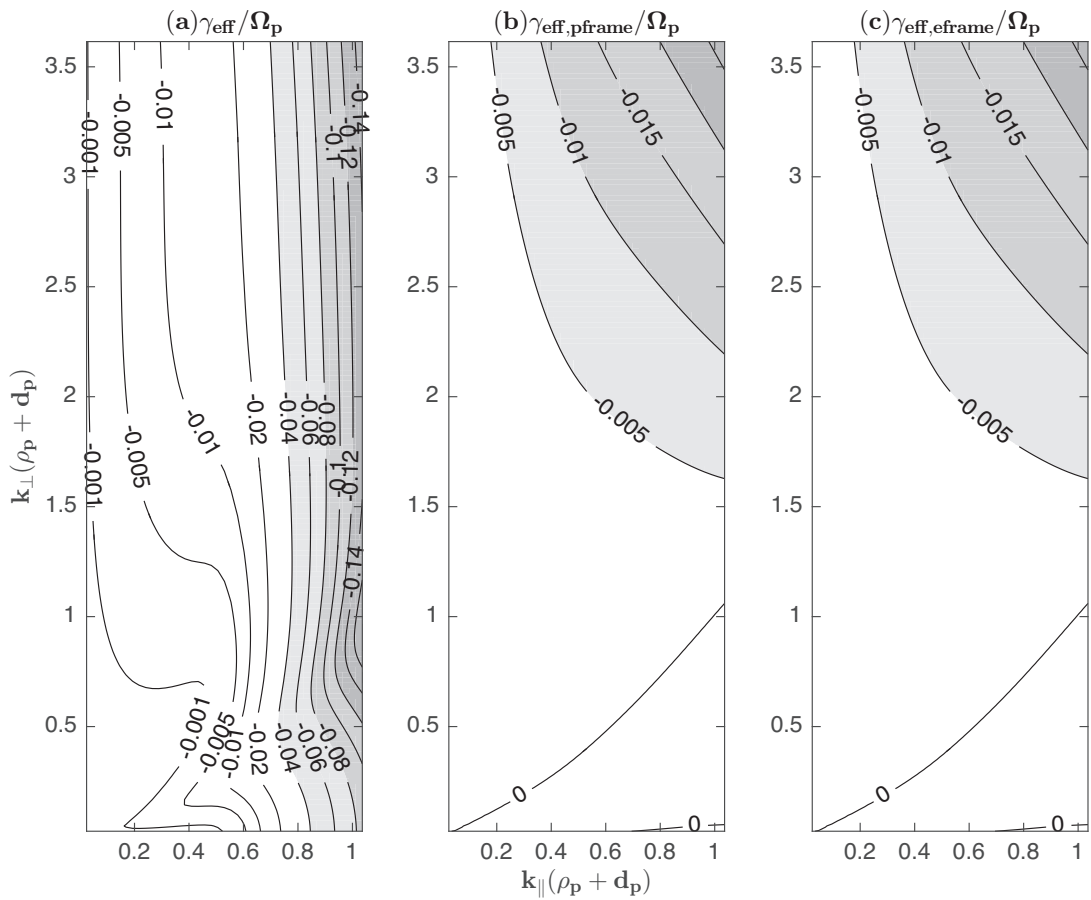}}
	\caption{Effective damping rate (conversion rate) calculated in different reference frames. (a) In the mean flow reference system (the same as 2(a)). (b) In the local flow reference frame of protons. (c) In the local flow reference frame of electrons. The damping rate calculated in the local frame is much smaller than the damping rate calculated in mean flow frame. Panels (b) and (c) are identical within numerical accuracy.}
	\label{fig3}
\end{figure}

In the local bulk flow reference frame, the distribution of the effective damping rate in wavenumber space is different from that in the mean bulk flow reference frame. Panels (b) and (c) in Figure \ref{fig3} show the effective damping rate in the local proton and electron bulk flow reference frame ($\mathbf{v}_{\rm{ref}} = \delta \mathbf{v}_{p}$ or $\delta\mathbf{v}_{e}$). These two panels are identical, because $\delta\mathbf{j}\cdot(\delta\mathbf{E}'_{p} - \delta\mathbf{E}'_{e}) = \delta\mathbf{j}\cdot[(\delta\mathbf{v}_{p} - \delta\mathbf{v}_{e})\times \mathbf{B}_0] = \delta\mathbf{j}\cdot(\delta\mathbf{j}\times \mathbf{B}_0) = 0$, where $\delta\mathbf{E}'_{p} $ and $\delta\mathbf{E}'_{e}$ are the fluctuating electric field in reference frames of $\mathbf{v}_{\rm{ref}} = \delta \mathbf{v}_{p}$ and $\mathbf{v}_{\rm{ref}} = \delta \mathbf{v}_{e}$, respectively. Fluctuating current and magnetic field do not change in the frame transformation, since we assume $\delta\mathbf{v}_p$ and $\delta\mathbf{v}_e$ to be much smaller than the speed of light. Compared to the mean bulk flow frame (panel (a)), the effective damping rate is much smaller. It means the $ \langle\delta\mathbf{j}\cdot \delta\mathbf{E}^\prime\rangle$ is much smaller in the local bulk flow frame, as shown by \citep{Wan2015}. The work done by the inertial force in the non-inertial frame (local bulk flow frame) is responsible for this imbalance. If we choose a periodically varying velocity as a reference velocity, the frame is by definition non-inertial.The work done by the resultant inertial force and its contribution to the energy transfer balance will be discussed in details in Section \ref{sec4}. Moreover, $\delta \mathbf{E}^\prime = \delta \mathbf{E} + \delta \mathbf{v}_{\rm{ref}} \times \mathbf{B}$ serves as a measure for the frozen-in condition. At large scales, waves follow the frozen-in condition, so that $\delta\mathbf{E}^\prime \approx 0$, and $\delta\mathbf{j}\cdot\delta\mathbf{E}^\prime \approx 0$ in the region of small $k$ (see Figure \ref{fig3}). At larger $k_\perp$ , however, $\delta\mathbf{E}' \ne 0$ and $\delta\mathbf{j}\cdot\delta\mathbf{E}^\prime \ne 0$, indicating frozen-in condition is thus broken at small scales as expected.

\section{Energy Distribution between Protons and Electrons}
\begin{figure}[htb!]
	\centerline{\includegraphics[width=16cm,clip=]{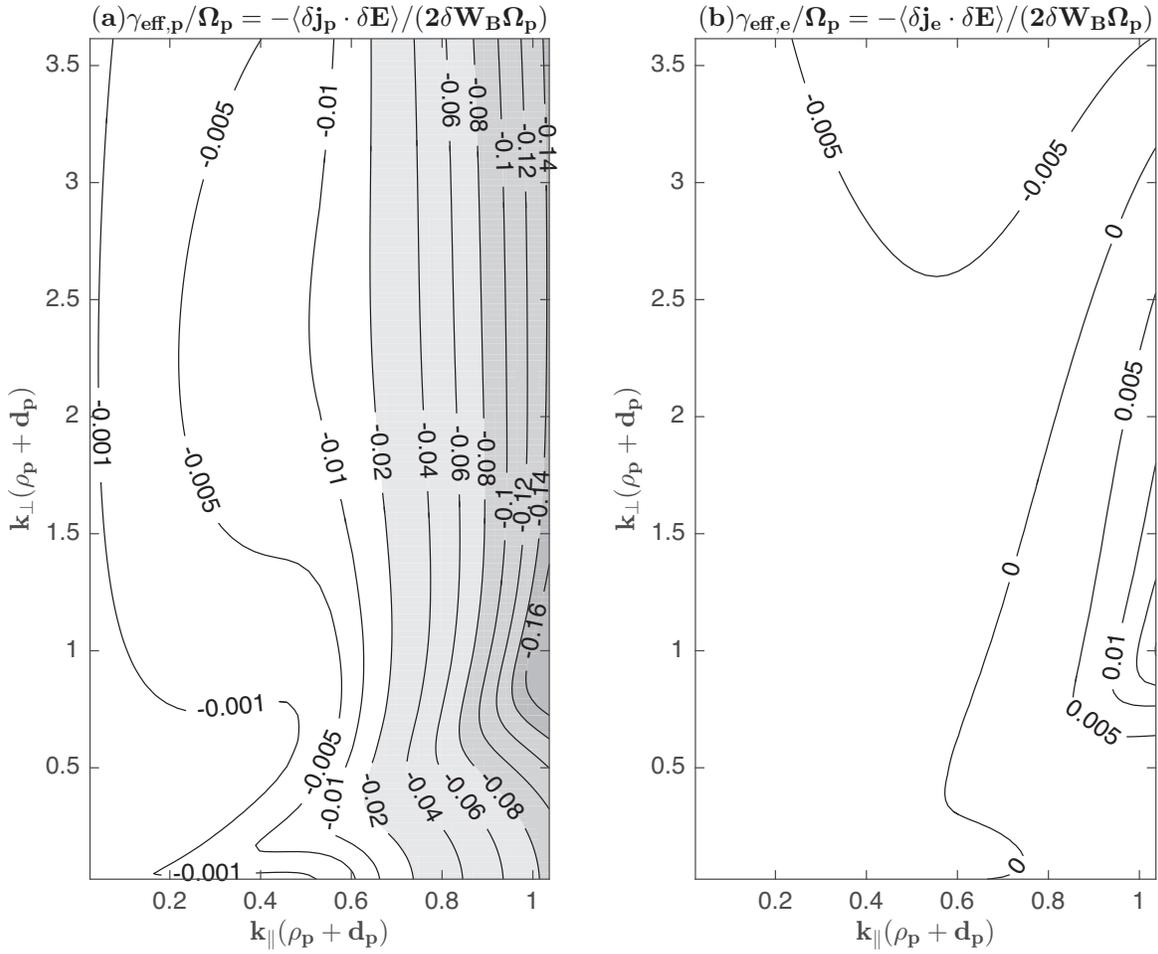}}
	\caption{Effective energy conversion rate (damping rate) for protons (a) and electrons (b). For protons the rate is negative, while the rate for electrons is positive at large $k_\parallel(\rho_p+d_p)$, suggesting that the electrons give energy to the EM fields at these scales.}
	\label{fig4}
\end{figure}

We now divide the fluctuating current into current populations carried by different species ($\delta\mathbf{j}_s = n_sq_s\delta \mathbf{v}_s $). The $\mathbf{j}_s\cdot\mathbf{E}$ term describes the amount of energy that is converted to protons and electrons separately. Figure \ref{fig4} shows the scale dependent effective damping rate $\gamma_{\rm{eff}}$for protons and electrons, respectively. Protons receive most of the EM energy, and their effective damping rate increases along $\mathbf{k}_\parallel$. Electrons receive almost no energy at small $\mathbf{k}$. In the range of $k_\parallel(\rho_p+d_p)\sim 1$, $\gamma_{\rm{eff,e}}/\Omega_p$ assumes a small and positive value, which means electrons transfer a small proportion of their kinetic energy to the EM fields. There are two possibilities why  $\langle\delta\mathbf{j}_s\cdot\delta\mathbf{E}\rangle$ can vanish. One is that the vectors $\delta\mathbf{j}$ and  $\delta\mathbf{E}$ are orthogonal to each other all the time, another is that the average of $\mathbf{j}_s\cdot\mathbf{E}$ over multiple periods is equal to zero . In the MHD Alfv\'enic range, the fluctuating electric field is perpendicular to the fluctuating velocity, so the effective damping rate in this region is zero. At smaller scales, however, kinetic effects introduce phase differences other than 90 degrees.

\begin{figure}[htb!]
	\centerline{\includegraphics[width=15cm,clip=]{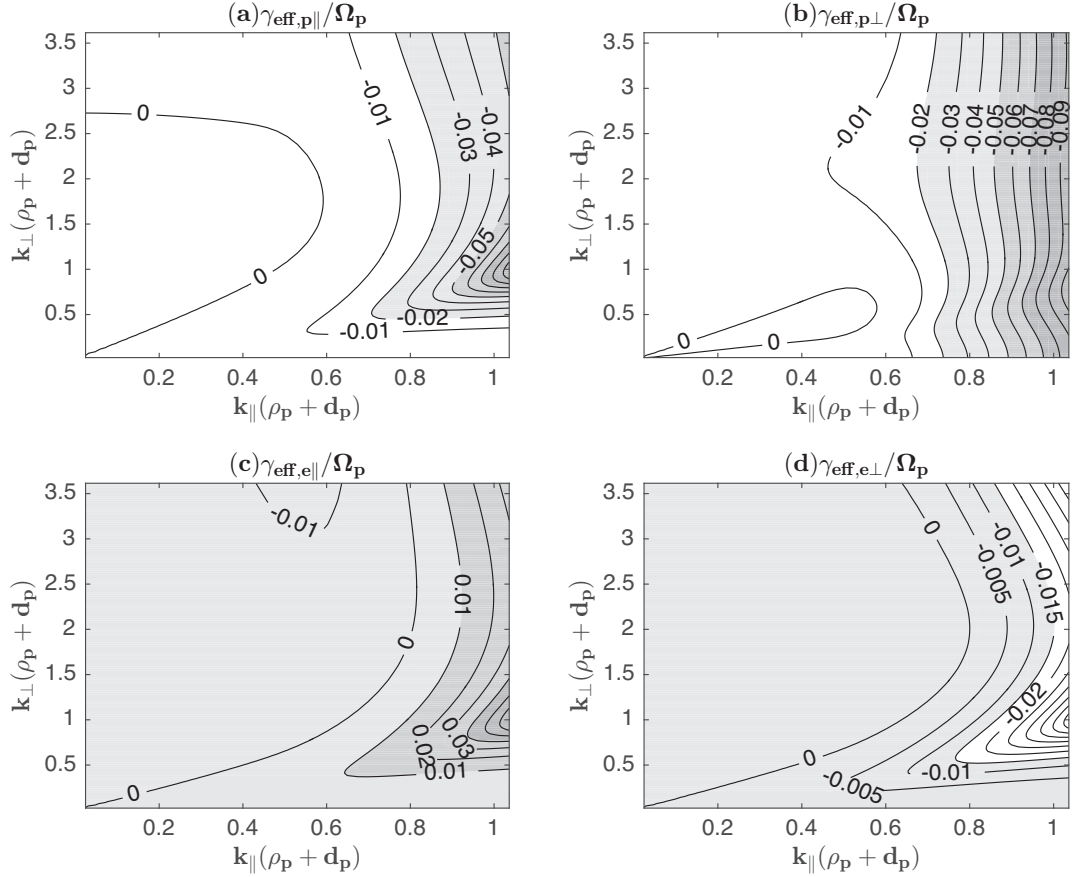}}
	\caption{Effective energy conversion rate (damping rate) for protons and electrons along parallel and perpendicular direction. (a) Conversion rate in parallel degrees of freedom for protons, (b) conversion rate in perpendicular degrees of freedom for protons, (c) conversion rate in parallel degrees of freedom for electrons, (d) conversion rate in perpendicular degrees of freedom for electrons.}
	\label{fig5}
\end{figure}

\begin{figure}[htb!]
	\centerline{\includegraphics[width=15cm,clip=]{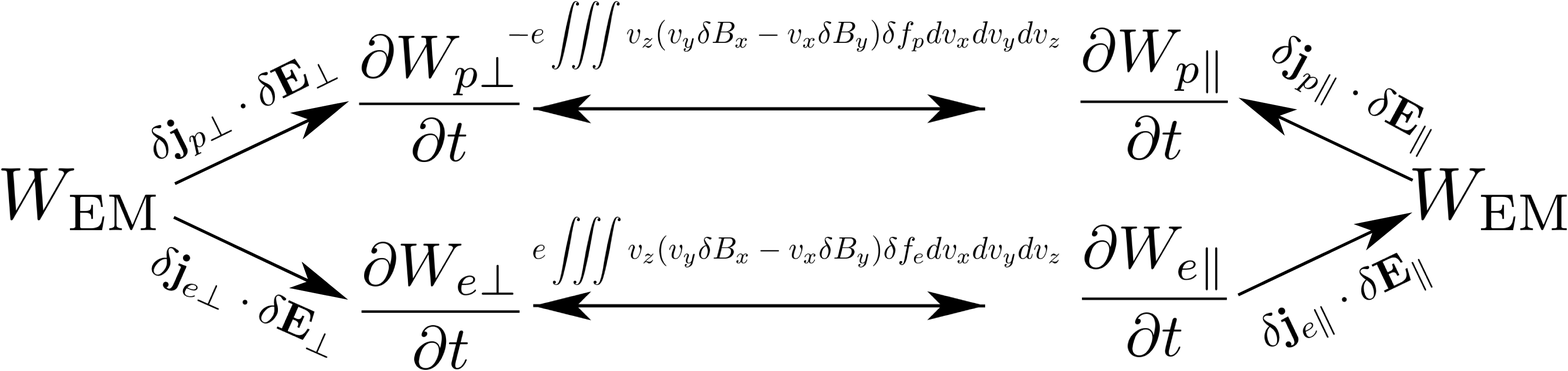}}
	\caption{The paths of energy conversion. EM field energy is transferred to particle kinetic energy via the electric field, and magnetic field fluctuations lead to a transfer between parallel and perpendicular degrees of freedom.}
	\label{fig6}
\end{figure}

As $\delta\mathbf{j}\cdot\delta\mathbf{E} = \delta\mathbf{j}_\parallel\cdot\delta\mathbf{E}_\parallel + \delta\mathbf{j}_\perp\cdot\delta\mathbf{E}_\perp$, the effective damping rates allow us to decompose the energy transfer between the parallel and perpendicular degrees of freedom. We show this separated distribution in wavenumber space in Figure \ref{fig5}. The protons gain more energy along the perpendicular direction than along the parallel direction, which is the result of cyclotron resonant wave-particle interactions. $\gamma_{\rm{eff},i\parallel/\Omega_p}<-0.01$ in the region ($k_\parallel(\rho_p+d_p)>0.6, k_\perp(\rho_p+d_p)>0.3$), which may be related to the energy transfer via Landau damping of KAWs along the parallel direction. The distributions of $\gamma_{\rm{eff},e\parallel/\Omega_p}$ and $\gamma_{\rm{eff},e\perp/\Omega_p}$ display an opposite pattern in wavenumber space. This opposite pattern suggests that particles are scattered in pitch-angle during the damping process. We quantify this effect by separating the kinetic-energy Equation \ref{eq1} into two kinetic-energy equations relating to the parallel and perpendicular kinetic energies as:
\begin{eqnarray}
	\frac{\partial \delta W_{s,\parallel}}{\partial t} +\nabla\cdot\delta\mathbf{Q}_\parallel&= \delta \mathbf{E}_\parallel\cdot\delta \mathbf{j}_{s,\parallel} + q_s\iiint{v_z(v_x\delta B_y - v_y\delta B_x)\delta f_s dv_xdv_ydv_z}\label{Eq5}
\\
	\frac{\partial \delta W_{s,\perp}}{\partial t} +\nabla\cdot\delta\mathbf{Q}_\perp&= \delta \mathbf{E}_\perp\cdot\delta \mathbf{j}_{s,\perp} + q_s\iiint{v_z(v_y\delta B_x - v_x\delta B_y)\delta f_s dv_xdv_ydv_z}\label{Eq6}
\end{eqnarray}
, where $\delta W_{s,\parallel} = m_s/2\iiint{v_z^2\delta f_s dv_xdv_ydv_z}$ and $\delta W_{s,\perp} = m_s/2\iiint{(v_x^2+v_y^2)\delta f_s dv_xdv_ydv_z}$ are the kinetic energies associated with the particle velocity in the parallel and perpendicular directions separately. Since we do not consider relative drifts in the mean-flow frame, particle kinetic energy is directly associated with thermal energy in our case. The LHS of Equations \ref{Eq5} and \ref{Eq6} represent the parallel and perpendicular energy transfer rates, which may be caused by Landau damping (parallel) and cyclotron damping (perpendicular) of electromagnetic energy. The Lorentz force leads to a transfer between the parallel and perpendicular degrees of freedom (RHS of Equation \ref{Eq5} and \ref{Eq6}), but it does not increase the total kinetic energy. Therefore, $\langle\delta\mathbf{E}_\parallel\cdot\delta\mathbf{j}_\parallel\rangle$ and  $\langle\delta\mathbf{E}_\perp\cdot\delta\mathbf{j}_\perp\rangle$ are not necessarily direct measures for Landau damping and cyclotron  damping under general conditions for waves with broad propagation angle. This scenario of energy transfer is shown in Figure \ref{fig6}. 

\section{Summary and Discussion} \label{sec4}
In this study, we compute the EM energy conversion rate spectra in wavenumber space. We define the effective damping rate $\gamma_{\rm{eff}}$, as the ratio of converted energy to magnetic field energy. Comparing the effective damping rate in the mean flow frame and local (oscillating) flow frame, we find that $\langle\delta\mathbf{j}\cdot\delta\mathbf{E}'\rangle$ does not appropriately reflect the transfer of energy between fields and particles, while $\langle\delta\mathbf{j}\cdot\delta\mathbf{E}\rangle$ is consistent with the damping rate of magnetic field energy. In the large $\mathbf{k}$ region around ion scales, most of the EM field energy is converted into proton kinetic energy rather than electron kinetic energy.

The energy partitioning between protons and electrons depends on various parameters, e.g., fluctuation amplitude, plasma $\beta$, and temperature ratio ($T_i/T_e$). Our study focuses on the ion scale under the typical solar-wind condition at 1 au. At smaller scales (electron scales), electrons receive more energy than protons via electron Landau damping of obliquely propagating KAWs \citep{Leamon1999}. Kinetic simulations show that the total heating rate of electrons increases relative to the heating rate of protons when both ion and electron kinetic scales are taken into account\citep{Matthaeus2016}. The application of our method to conditions with low ion plasma $\beta_i$  ($\beta_i=0.1$) lead to results at ion scales that are similar to our results presented for ion plasma $\beta_i$. These results for low plasma beta can help to understand the PSP measurements in the inner heliosphere in a future project; however, a detailed study of these conditions is beyond the scope of this work.

In addition, the $\langle\delta\mathbf{j}\cdot\delta\mathbf{E}\rangle$ term only describes the conversion between EM-field energy and particle kinetic energy, and does not provide information about the conversion between the bulk kinetic energy and thermal kinetic energy. When transformed between different reference frames, the velocity distribution function just shifts as a whole in velocity space. Both $\delta\mathbf{E}$ and $\delta\mathbf{E}'$ work on all of the particles, so they only contribute to the energy transfer into bulk kinetic energy. There is no direct energy transfer from EM energy to thermal energy. However, for the dissipation of Alfv\'enic turbulence, both fluctuating EM-field energy and fluctuating bulk kinetic energy will eventually be dissipated and converted into thermal kinetic energy. 

The power of the inertial force for a particle species depends on the amplitude of the velocity fluctuation, the wave frequency, and the mass of a particle of the given species: $P_{iner,s}=-m_s(d\delta \mathbf{v}_s/dt)\cdot\delta \mathbf{v}_s$. The effect of the inertial force becomes more significant at smaller scales, as $\omega$ increases with decreasing scale. Its effect on electrons may be neglected compared to that on protons because the electron mass is much smaller than the proton mass.

There is a possible way to include $\delta\mathbf{j}_s\cdot\delta\mathbf{E}'$ in the governing equation for thermal kinetic energy. Substituting $\mathbf{E}'=\mathbf{E}+\mathbf{v}_{bs}\times\mathbf{B}$ into the momentum equation ($\mathbf{v}_{bs}$ is the bulk flow velocity of species s), leads to
\begin{eqnarray}
n_sm_s\frac{d\mathbf{v}_{bs}}{dt} = n_sq_s(\mathbf{E}+\mathbf{v}_{bs}\times\mathbf{B})-\nabla\cdot\mathbf{P}_s&=n_sq_s\mathbf{E}'-\nabla\cdot\mathbf{P}_s
\end{eqnarray}
Multiplying the equation with $\mathbf{v}_{bs}$ yields
\begin{eqnarray}
\mathbf{v}_{bs}\cdot(\nabla\cdot\mathbf{P}_s) = n_sq_s\mathbf{v}_{bs}\cdot\mathbf{E}'-n_sm_s\mathbf{v}_{bs}\cdot\dfrac{d\mathbf{v}_{bs}}{dt}&=\mathbf{j}_s\cdot\mathbf{E}' - n_sm_s\mathbf{v}_{bs}\cdot\dfrac{d\mathbf{v}_{bs}}{dt}
\end{eqnarray}
The term $n_sm_s\mathbf{v}_{bs}\cdot(d\mathbf{v}_{bs}/dt)$ is the change rate of bulk kinetic energy. In the non-inertial frame, this is the power due to the inertial force to guarantee energy conservation. Substituting this term into the thermal energy equation ($W_{th,s} = m_s/2\iiint{(\mathbf{v}-\mathbf{v}_{bs})^2f_sdv_xdv_ydv_z}$ is the particle thermal energy) leads to
\begin{eqnarray}
\frac{\partial W_{th,s}}{\partial t} + \nabla \cdot (W_{th,s}\mathbf{v}_{bs} + \mathbf{h}_s + \mathbf{P}_s\cdot\mathbf{v}_{bs}) &= \mathbf{j}_s\cdot\mathbf{E}' - n_sm_s\mathbf{v}_{bs}\cdot\dfrac{d\mathbf{v}_{bs}}{dt}, \label{Eq9}
\end{eqnarray}
 where $\mathbf{h}_s = m_s/2\iiint{(\mathbf{v}-\mathbf{v}_{bs})^2(\mathbf{v}-\mathbf{v}_{bs})f_sdv_xdv_ydv_z}$ is the heat flux vector. Note that in association with the appearance of $\mathbf{j}_s\cdot\mathbf{E}'$ ,the power associated with the inertial force also exists in Equation \ref{Eq9}. These derivations show that $\mathbf{j}\cdot\mathbf{E}'$ cannot fully describe the energy transfer to kinetic thermal energy, except if $n_sm_s\mathbf{v}_{bs}\cdot(d\mathbf{v}_{bs}/dt)=0$. The combination of $\mathbf{j}\cdot\mathbf{E}'$ and $n_sm_s\mathbf{v}_{bs}\cdot(d\mathbf{v}_{bs}/dt)$, which is the same as $\mathbf{v}_{bs}\cdot(\nabla\cdot\mathbf{P}_s)$, must be taken into consideration. Whether the energy transfer between EM fields and particles($\sum_s\mathbf{j}_s\cdot\mathbf{E}_s$) is less or greater than the energy transfer between bulk kinetic energy and thermal kinetic energy ($\sum_s (\mathbf{P}_s\cdot\nabla)\cdot \mathbf{v}_{bs}$) is another interesting question to be addressed in the future through theoretical calculation and observational analysis. \citet{Yang2019} found that the scale-dependent $-(\mathbf{P}_s\cdot\nabla)\cdot \mathbf{v}_{bs}$ dominates the energy conversion at smaller scales in their 2.5D kinetic simulations.
 
The exact contributions of Landau and cyclotron resonances are difficult to estimate. For example, $\langle\mathbf{j}_\perp\cdot\mathbf{E}_\perp\rangle$ represents the total rate of energy conversion in the perpendicular direction, including the contributions from the particles satisfying the cyclotron-resonance condition and other particles outside the resonant velocity range, as long as they carry part of the current $\mathbf{j}_\perp$. On the other hand, the particle scattering in the phase space due to cyclotron resonance is also governed by the Lorentz force of the fluctuating magnetic field, which transfers energy between perpendicular and parallel degrees of freedom, and acts together with the electric force to form the diffusion plateau of cyclotron resonance in phase space. For $\langle\mathbf{j}_\parallel\cdot\mathbf{E}_\parallel\rangle$, the situation is similarly consisting of both the Landau resonance part and the non-resonant part. Like in the cyclotron-resonant case, its strength depends on the distribution of the particle-phase-space density. At small $\theta_{kB}$ (the angle between $\mathbf{B}_0$ and $\mathbf{k}$) and large scales, the effect of ion-cyclotron resonances is presumably stronger because the resonance condition is easier to satisfy. At larger $\theta_{kB}$ and smaller scales, Landau damping may plays more important role \citep{Leamon1999}.
 
Our results show a significant energy transfer around the scale $k_\parallel(\rho_i+d_i)\sim 1$. This scale is related to the proton cyclotron resonance, which may lead to the spectral break observed in the magnetic-field power spectra in solar-wind turbulence \citep{Duan2018, Duan2020}. The spectral break may also be caused by the transition of Alfv\'enic turbulence to dispersive Alfv\'enic turbulence around the ion scale. Future work is planned to compute the energy conversion rate spectrum based on in-situ measurements in space, and investigate its relation to the mechanisms responsible for the spectral break. The radial evolution of spectral break in the inner heliosphere and its underlying physical processes of diffusion, dissipation, and dispersion in the evolving solar wind streams will be one of the key issues when investigating the solar wind turbulence measurements from PSP \citep{He2019c}.

\bigbreak

\noindent Acknowledgements:

This work at Peking University is supported by NSFC under contracts 41574168, 41674171, 41874200, and 41421003,  and also supported by the project of Civil Aerospace "13th Five Year Plan" Preliminary Research in Space Science with Project Number of D020301. D.V. is supported by the STFC Ernest Rutherford Fellowship ST/P003826/1 and STFC Consolidated Grant ST/S000240/1. 

\bibliographystyle{aasjournal}
\bibliography{reference}

\end{document}